Title: First detections of the [NII] 122 μm line at high redshift: Demonstrating the utility of the line for studying galaxies in the early universe.

Running Title: First detection of the [NII] 122 μm line

Carl Ferkinhoff[1]
Drew Brisbin[1]
Thomas Nikola[1]
Stephen C. Parshley[1]
Gordon J. Stacey[1]
Thomas G. Phillips[2]
Edith Falgarone[3]
Dominic J. Benford[4]
Johannes G. Staguhn[4,5]
Carol E. Tucker[6]

[1] Department of Astronomy, Cornell University, Ithaca, NY 14853, USA; cferkinh@astro.cornell.edu
[2] California Institute of Technology, Pasadena, CA 91125, USA
[3] LERMA, CNRS, Observatoire de Paris and ENS, France
[4] Observational Cosmology Laboratory (Code 665), NASA Goddard Space Flight Center, Greenbelt, MD 20771, USA
[5] Department of Physics & Astronomy, Johns Hopkins University, Baltimore, MD 21218, USA.
[6] Department of Physics and Astronomy, Cardiff University, Cardiff CF24 3AA, UK.

We report the first detections of the [NII] 122 μm line from a high redshift galaxy. The line was strongly (> 6σ) detected from SMMJ02399-0136, and H1413+117 (the Cloverleaf QSO) using the Redshift(z) and Early Universe Spectrometer (ZEUS) on the CSO. The lines from both sources are quite bright with line-to-FIR continuum luminosity ratios that are ~7.0×10$^{-4}$ (Cloverleaf) and 2.1×10$^{-3}$ (SMMJ02399). With ratios 2-10 times larger than the average value for nearby galaxies, neither source exhibits the line-to-continuum deficits seen in nearby sources. The line strengths also indicate large ionized gas fractions, ~8 to 17% of the molecular gas mass. The [OIII]/[NII] line ratio is very sensitive to the effective temperature of ionizing stars and the ionization parameter for emission arising in the narrow-line region (NLR) of an AGN. Using our previous detection of the [OIII] 88 μm line, the [OIII]/[NII] line ratio for SMMJ02399-0136 indicates the dominant source of the line emission is either stellar HII regions ionized by O9.5 stars, or the NLR of the AGN with ionization parameter log(U) = -3.3 to -4.0. A composite system, where 30 to 50% of the FIR lines arise in the NLR also matches the data. The Cloverleaf is best modeled by a superposition of ~200 M82 like starbursts accounting for all of the FIR emission and 43% of the [NII] line. The remainder may come from the NLR. This work demonstrates the utility of the [NII] and [OIII] lines in constraining properties of the ionized medium.

Subject Headings: galaxies: individual (SMM J02399-0136, H1413+117) galaxies: high-redshift – galaxies: starburst – galaxies: active – submillimeter: galaxies

# 1. INTRODUCTION

The far-infrared (FIR) fine-structure lines from carbon, nitrogen, oxygen and their ions serve as important and in some cases dominant coolants of major phases of the interstellar medium (ISM). In the 80's, 90's, and first decade of the new century much work was done characterizing these lines in the Milky Way and nearby galaxies (e.g Stacey et al. 1991, Wright et al. 1991, Malhotra et al. 2001, Luhman et al. 2003, Brauher et al. 2008, and Fisher et al. 2010). The advent of a new generation of sensitive ground and spaced based submillimeter instruments (e.g. ZEUS, Z-Spec and Herschel-SPIRE) is now enabling the study of these lines to move beyond the local Universe. Lines like the [CII] 158 μm, [OIII] 88 and 52 μm, and now the [NII] 122 μm line, redshifted into the submillimeter bands, are providing important details on the physical conditions and excitation of gas in these early galaxies (e.g. Maiolino et al. 2005, Maiolino et al. 2009, Hailey-Dunsheath et al. 2010, Sturm et al. 2010, Ivison et al. 2010a, Stacey et al. 2010, Ferkinhoff et al. 2010, and Valtchanov et al. 2011), essential to our understanding of the star formation process during the peak epoch of star formation, z~1 to 3 (Hopkins & Beacom 2006). Here we report the detection of the [NII] 122 μm line from SMMJ02399-0136 (hereafter SMMJ02399) at z=2.81 and H1413+117(the Cloverleaf QSO) at z=2.56. These detections are the first reported detection of this line from a galaxy at z>0.09.

Nitrogen requires 14.5 eV photons for single ionization, so that $N^+$ resides predominantly in HII regions formed by late O or early B type stars. The ground state term of $N^+$ is split into three levels leading to the 122 μm ($^3P_2 - ^3P_1$), and 205 μm ($^3P_1 - ^3P_0$) fine structure lines. Because the lines are collisionally populated, optically thin, and insensitive to gas temperature their line intensity yields a minimum ionized-gas mass and the minimum numbers of photons capable of maintaining the ionization equilibrium for the species. For the [NII] lines this means photons between 14.5 < hν < 30 eV and spectral types between B2 and O8 stars[7].

Comparing the line flux of [NII] to [OIII] provides a very sensitive probe of UV field hardness since 35 eV photons are needed to form $O^{++}$. Furthermore, because the [OIII] 88 μm and [NII] 122 μm emitting levels have similar critical densities for thermalization (510 $cm^{-3}$, and 310 $cm^{-3}$ respectively), the line ratio is very insensitive to gas density. This means that for stellar radiation fields the line ratio constrains the most luminous star on the main sequence and hence the age of the stellar population (Figure 1, bottom axis). These lines also arise from the narrow line region (NLR) of active galactic nuclei (AGN), in which case the ratio is a sensitive indicator of the ionization parameter, U, of the NLR—defined by the ratio of ionizing photons to number density of the gas (Abel et al. 2009).

SMMJ02399 was the first submillimeter-selected galaxy (Ivison et al 1998). Lensed by a factor μ~2.38 (Ivison et al 2010b, hereafter I10b), it still has a very large intrinsic FIR luminosity[8], $L_{FIR}$~1.22×$10^{13}$ $L_\odot$ (Frayer et al 1998) and molecular gas content, M($H_2$)~1×$10^{11}$ $M_\odot$ (I10b). SMMJ02399 is a composite AGN/starburst system that contains 4 distinct components within ~3" radius in the sky-plane. Initial BVR imaging identified two components: L1, associated with the broad-absorption-line quasar (BAL); and L2, which exhibits very blue emission extending ~3" to the east of L1 (Ivison et al 1998). I10b identified 2 additional components—L1N and L2SW—in HST/ACS/NICMOS images and argue that the FIR luminosity arises from L2SW based on VLA 1.4 GHz continuum and EVLA CO(1-0) maps. While the FIR emission is dominated by L2SW, the combined SED of the entire system is dominated by the AGN in L1. Because L2SW has very red near/mid-infrared colors, and is coincident with the molecular gas and 1.3mm continuum emission peaks, it is likely the site of a massive young starburst reflected in its large FIR luminosity (I10b). It is suggested that the individual components may be independent galaxies at different stages of a merger process (I10b). With its 11" FWHM beam, the present observations using ZEUS/CSO enclose all of the components.

---

[7] Stellar-types from Vacca et al. 1996
[8] We assume a flat cosmology with $\Omega_\Lambda$= 0.73 and $H_0$=71 km $s^{-1}$ $Mpc^{-1}$

In Ferkinhoff et al. 2010 (hereafter F10) detected the [OIII] 88 μm line from SMMJ02399 with $L_{[OIII]}/L_{FIR} \sim 3.6 \times 10^{-3}$. By comparing with the Hα and [OIII] 5007 lines, they argue that the system contains a massive starburst with a stellar mass-distribution topped by O7.5 stars and an ionized gas mass of $M_{HII} \sim (1-10) \times 10^9\ M_\odot$. Our present detection of the [NII] 122 μm line refines and constrains this estimate.

The Cloverleaf quasar (H1413+117) is also a lensed (μ~11) BAL quasar. First detected in an optical spectroscopic survey (Hazard et al. 1984), it is a composite system with a partially obscured AGN dominating the $\sim 7 \times 10^{13}\ L_\odot$ bolometric luminosity. Based on strong PAH emission, Lutz et al. (2007) conclude that an extreme starburst dominates the FIR, with $L_{FIR} \sim 5$-10% $L_{Bol}$. The Cloverleaf is well studied in molecular gas tracers (e.g. CO, [CI], $^{13}$CO; c.f., Weiß et al 2003, Barvainis et al 1997, Bradford et al. 2009, Henkel et al. 2010). Bradford et al. (2009) find that its molecular gas reservoir, $M(H_2)=0.2-5\times10^{10}\ M_\odot$, fills a disk between 325 and 650 pc in radius and the hard X-ray flux from the AGN heats the molecular ISM, resulting in a top heavy stellar IMF.

**2. OBSERVATIONS**

We used the Redshift(z) and Early Universe Spectrometer (ZEUS; Hailey-Dunsheath 2009, Stacey et al. 2007) at the 10.4 meter Caltech Submillimeter Observatory (CSO)[9] on Mauna Kea, Hawaii. ZEUS is a direct-detection echelle-grating spectrometer with R~1000 (see above references for details). Both sources were observed in January 2011 in a chopping and nodding mode with a chop rate of 2 Hz and azimuthal amplitude of 30". For SMMJ02399, the line was detected (6.4σ) in 1.9 hours integration time through a line of sight (LOS) transmission of 32%. For the Cloverleaf, the line was detected (5.7σ) in 1.5 hours with a LOS transmission of 42%. Spectral calibration was obtained using a CO-absorption gas cell while flux calibration and spectral response flats were obtained with a $LN_2$ cold load. Telescope coupling was measured with observations of Uranus. The final spectra are shown in Figure 2.

## 3. RESULTS

### 3.1 Line luminosity

Table 1 lists the source and line parameters. The line is bright with $L_{[NII]}/L_{FIR}\sim7.0\times10^{-4}$ and $2.1\times10^{-3}$ for the Cloverleaf and SMMJ02399, respectively. For nearby systems the typical range is $L_{[NII]}/L_{FIR}\sim3\times10^{-5}$ to $1\times10^{-3}$ (Graciá-Carpio et al. 2010). While the Cloverleaf does lie within this range, it is ~2 times larger than the average ratio. SMMJ02399, on the other hand, is 10 times the average value. The lines from both sources are clearly narrow (<400-600 km/s) suggesting the line arises from the NLR and/or from HII regions.

---

[9] The CSO telescope is operated by the California Institute of Technology, under funding from the NSF, grant AST-0838261.

Genzel et al. (2003) detect two velocity components in the CO emission from SMMJ02399 at ~ 420 km/s to the red and blue of the nominal systemic redshift of z=2.8076 that are associated with the L1 and L2 components respectively. F10 only detected one component in the [OIII] 88 μm line, which at low significance, is shifted ~400 km/s to the blue of z=2.8076. Here we clearly detect the blue component in the [NII] 122 μm line with a hint (2.9σ) of the red component (1.54±0.54×10$^{-18}$ W m$^{-2}$). All values and discussion are restricted to the blue, L2 associated component of the [NII] line. Subsequently, we strongly prefer the star-formation explanation of the line emission even though we will show (section 4.2.1) that both the [NII] and [OIII] line fluxes are consistent with having arisen in the NLR.

## 3.2 Minimum Mass of Ionized Gas

A minimum ionized gas mass, $M_{min}(H^+)$, associated with the line emission occurs in the high-density, high-temperature limit assuming that all nitrogen in the HII regions is singly ionized (see F10). This occurs in regions ionized by B2 to O8 stars. $M_{min}(H^+)$ is then:

$$M_{min}(H^+) = F([NII]) \frac{4\pi D_L^2}{\frac{g_2}{g_t} A_{21} h\nu_{21}} \frac{m_H}{\chi(N^+)},$$

Where $A_{21}$ and $g_2$ are the Einstein A coefficient (7.5×10$^{-6}$ s$^{-1}$) and statistical weight (5) of the $^3P_2$ emitting level, $g_t = \sum g_i \exp(-\Delta E_i/kT)$ is the partition function, $h$ is Planck's constant, $\nu_{21}$ is the 122 μm line frequency, $D_L$ is the luminosity distance, $m_H$ is the hydrogen mass, and $\chi(N^+)$ is the $N^+/H^+$ abundance ratio. For the minimum mass case, $\chi(N^+)=\chi(N)$, the total N/H abundance ratio. We assume "HII region" gas-phase nitrogen abundance for both sources (9.3×10$^{-5}$, Savage & Sembach 1996), so that $M_{min}(H^+)$ = 4.0×10$^{10}$ $M_\odot$ and 2.5×10$^9$ $M_\odot$, and the minimum ionized-gas to molecular-gas mass ratio, $M_{min}(H^+)/M(H_2)$, is 0.17 and 0.08 for SMMJ02399 and the

Cloverleaf, respectively. For SMMJ0299 this is ~10 times larger than the ratio obtained by F10 based on the [OIII] 88 μm line. However, if the HII regions are formed by cooler stars than the F10 model, with effective temperatures of ~36,000 K for example, then only ~13% of the oxygen is doubly ionized and the minimum ionized-gas/molecular-gas mass ratio obtained using the [OIII] line is consistent with the value obtained using the [NII] line above.

## 4. DISCUSSION

### 4.1 Ionized gas mass

It is clear that both SMMJ02399 and the Cloverleaf have a very significant amount of ionized gas. Figure 3 shows $M_{min}(H^+)/M(H_2)$ versus the [OIII] 88 μm/[NII] 122 μm line ratio for SMMJ02399, the high-z source SMMJ2135-0102 (Ivison et al. 2010a), and nearby sources observed by ISO (Brauher et al. 2008), all calculated using the method described above. The line ratio for our source is similar to that of the other sources, but the *minimum* ionized mass fractions for the Cloverleaf (8%) and SMMJ02399 (17%), are significantly larger than for the nearby sources. However, we estimate $M_{min}(H^+)/M(H_2)$~1% for M82, while Lord et al. (1996) determined $M_{min}(H^+)/M(H_2)$~12% for this nearby starburst galaxy. This suggests that $M_{min}(H^+)/M(H_2)$ for many local galaxies may be significantly larger than the minimum values in Figure 3, so that the high-z mass-fractions may not be unusual.

The very large star-formation rates observed for both SMMJ02399 and the Cloverleaf might naturally produce their large $M_{min}(H^+)/M(H_2)$ fractions. A higher SFR means more HII regions and more ionized gas to produce the bright [NII] emission we observe. To test this idea we plot (Fig. 3, right) the SFR surface density versus $M_{min}(H^+)/M(H_2)$ for the nearby sample (only the

[NII] derived values are shown), SMMJ02399, the Cloverleaf, APM08279+5255 (based on [OIII] from F10), and a set of high-z [CII] sources (Stacey et al. 2010), where $M_{min}(H^+)/M(H_2)$ is estimated by assuming 30% of the observed [CII] line emission arises in HII regions (Oberst et al. 2006). There is indeed a strong trend of increasing $M_{min}(H^+)/M(H_2)$ with increasing star formation. Additionally, all of the high-z sources with SFR$\gtrsim$1000 $M_\odot yr^{-1}$ reside at the upper end of the star-formation/mass-fraction relation suggesting that the large ionized mass can be accounted for solely by their enhanced SFR.

Graciá-Carpio et al. (2011) find a FIR-line/FIR-continuum deficit for all FIR lines observed with Herschel/PACs with $L_{FIR}/M(H_2) \gtrsim 80$ $L_\odot M_\odot^{-1}$. They argue that the deficit results from higher ionization parameters in these systems and the line deficits and the higher ionization parameters are signatures of merger driven star formation with an associated increase in star formation efficiency. Our observations, those from F10, and the recent [OIII] 88 μm detection from H-ATLAS J090311.6+003906 (Valtchanov et al. 2011) show no such deficit. At present, there are few high-z observations available to test for a high-z deficit, but if future observations show similar ratios to those reported here, then the Garcia-Carpio et al. (2011) model does not describe the high-z population.

**4.2 Gas Excitation Mechanisms**

Both SMMJ02399 and the Cloverleaf are composite systems, so that [NII] emission may arise in the AGN's NLR and/or stellar HII regions. Unfortunately only a few UV, optical or IR lines are reported so that standard AGN/star-forming diagnostics are of limited utility (e.g. BPT diagram; Kewley et al. 2006). Here we analyze the [NII] 122 μm line in both star-forming and AGN paradigms using the HII region models from Rubin (1985) and the NLR models of Groves et al. (2004).

Rubin (1985) calculates the expected intensities of HII-region emission lines as functions of gas density, effective stellar temperature, $T_{eff}$, and metallicity. Here we use the "K" models with O/H=$6.76\times10^{-4}$ and N/H=$1.15\times10^{-4}$. These abundances are near the "HII region" values of Savage & Sembach (1996). We also examine the lower metallicity "D" models with O/H=$1.27\times10^{-4}$, and N/H=$1.42\times10^{-5}$. In all cases we use the "49" models with stars producing $10^{49}$ Lyman continuum photons per second regardless of $T_{eff}$.

The Groves et al. (2004) models are radiation-pressure dominated and include dust. Their parameter space includes typical ranges for gas density ($n_H = 10^2$ to $10^4$ cm$^{-3}$), power-law index of the ionizing source ($\alpha = -1.2$ to $-2.0$), and ionization parameter, $log(U) = 0$ to $-4.0$. We restrict our analysis to the solar-metallicity, $1Z_\odot$, and 25% solar-metallicity, $0.25Z_\odot$, models with abundances similar to the "K" and "D" Rubin (1985) models.

**4.2.1 SMMJ02399**

For SMMJ02399, F10 compared their [OIII] 88 μm line to sparsely sampled [OIII] 5007Å and Hα maps, and found the gas is ionized by O7.5 ($T_{eff}$=40,000) stars and has $n_e$ ~100-1000 cm$^{-3}$, depending on metallicity. Here, using only the [OIII] 88 μm/[NII] 122 μm line ratio, the best fit HII region models have similar densities: 100 ("K" model) or 1000 cm$^{-3}$ ("D" model). However, the line ratio strongly constrains the ionization source at O9.5 ($T_{eff}$=35,000 K) stars ionizing a total of $(1-4)\times10^9$ HII regions. These stars are significantly cooler than expected by F10 demonstrating the utility of the two lines to tightly constrain the stellar populations.

The best fit "K" model suggests an ionized gas mass of $4.8 \times 10^{10}\,M_\odot$ or 16% of the molecular gas mass, in agreement with the estimate in Section 4.2. The lower-metallicity "D" models predict $M_{min}(H^+)/M(H_2) \sim 48\%$. The increase in $M_{min}(H^+)$ reflects the lowered N and O abundance. The total luminosity of the stars from the "K" model solutions accounts for 100% of $L_{FIR}$, but is 28 times larger for the "D" model. This eliminates the "D" model solution since detailed studies of SMMJ02399 require that $L_{optical} \leq L_{FIR}$ (Ivison et al. 1998). Curiously, the optical line estimates from F10 do not fit the present models. For the best fit "K" HII region model, the [OIII] 5007 and Hα line predictions are respectively 10 times weaker, and 4 times stronger than observed. However, the discrepancy between the optical and FIR lines is not entirely unexpected since the optical lines are very susceptible to extinction effects and their strengths were scaling estimates (F10).

For the NLR models the [NII] 122 μm/[OIII] 88 μm line ratio constrains ionization parameter to between log(U) = -3.4 to -3.7 for the solar-metallicity model (see Figure 1) and *log(U) = -3.3 to -4* for the low metallicity model (Groves et al. 2004). The power-law index is also slightly constrained for the low-metallicity model to between -2 and -1.7. The [OIII] 5007 Å/ 88 μm line ratio provides a constraint on gas density, yielding $n_e \sim 500$ cm$^{-3}$ and $n_e \sim 500\text{-}1000$ cm$^{-3}$ for the solar and quarter-solar metallicity models respectively. With the lines observed, we are unable to constrain the metallicity. The observed [OIII] 5007 Å /Hα ratio does not fit our NLR model as the Hα fluxes are 10 times weaker than the NLR model predicts.

The fact that the [OIII] 5007Å strength is inconsistent with our HII region model, but consistent with our NLR model solutions suggests a composite model is appropriate. Indeed the AGN is likely responsible for between 25 to 75% of the FIR luminosity (Frayer et al. 1998, Bautz et al.

2000). A composite solution has HII regions, with $n_e$~100 cm$^{-3}$, formed by slightly coolers stars (O9.5, $T_{eff}$=34,500 K) than the pure star-formation model. The NLRs can then account for half the line luminosity given a slightly higher ionization parameter, $log(U)$ = -3.3 to -3.45, and $n_e$~500 cm$^{-3}$. Alternatively, for an HII region with $n_e$~1000 cm$^{-3}$ and a NLR ionization parameter, $log(U)$ = -3.25 to -3.4, ~30% of the line emission is attributable to the AGN and ~70% to star-formation.

**4.2.2 The Cloverleaf (H1413+117)**

Besides our [NII] 122 μm detection, there are no other FIR lines detected from the Cloverleaf to for constraining source properties. However, the 6.2 and 7.7 μm PAH features (Lutz et al 2007), and the Hα (1.3×10$^{-20}$ W cm$^{-2}$), Hβ (3.3×10$^{-21}$ W cm$^{-2}$), and the [OIII] 5007Å (1.0×10$^{-21}$ W cm$^{-2}$) lines are detected (Hill et al. 1993). Due to the moderate resolving power of the optical observations it is not clear if any emission arises from broad-line components. Here we assume that the optical lines are narrow and arise either in the NLR or stellar HII regions.

The [OIII] 5007Å/Hα and [NII] 122 μm/Hβ ratios and line strengths are best fit by Rubin (1985) "K" models with (2-3)×10$^7$ HII regions ionized by O9–O8.5 stars ($T_{eff}$=36,000–37,000 K). These HII regions have $n_e$~100 cm$^{-3}$, and $M_{HII}$~(8–10)×10$^9$ M$_\odot$ (the lower value for the O9 stars), so that $M_{HII}/M(H_2)$~30%. The line ratios are inconsistent with the low metallicity "D" models. The best fit HII region model can account for 50 to 80% of the observed $L_{FIR}$.

This HII region model is similar to that obtained for M82 (Lord et al. 1996, Colbert et al. 1999). However, Weiß et al. (2003) determine that the Cloverleaf contains both a 115K warm-dust component—likely heated by the AGN—and a 50K cool-dust component that Lutz et al. (2007)

attribute to a massive starburst, which dominates the FIR luminosity of the system. The Cloverleaf's cool-dust component is well matched by the single dust component in M82 (48 K, Colbert et al 1999; 50K, Negishi et al. 2001) as are the Cloverleaf's 6.2 and 7.7 µm PAH feature to FIR continuum ratios (Table 2). Therefore we model the Cloverleaf starburst as a superposition of ~200 M82-like starbursts to account for the Cloverleaf's FIR luminosity and molecular-gas mass being ~200 times larger than M82's. However, the Cloverleaf's [NII] line is ~470 times stronger than M82's, consequently this model can only attribute ~43% of the observed [NII] luminosity to star formation and the associated the cool-dust component of the Cloverleaf. The remaining [NII] line flux may arise, along with the warm-dust component, in the NLR of the Cloverleaf.

## 5. SUMMARY AND OUTLOOK

We have made the first detections of the [NII] 122 µm line at high-z. The line traces AGN and starburst activity providing constraints on stellar populations and AGN/NLR ionization parameters. Further constraints on source properties require additional FIR spectroscopy and the high spatial resolution that Herschel and/or ALMA may provide.

We find that high-z galaxies with high-SFR have large quantities of ionized gas resulting in strong line emission. At high redshifts, there is no the decline in $L_{FIR-Line}/L_{FIR}$ at high $L_{FIR}$ as seen locally (Graciá-Carpio et al. 2011), so that the FIR lines may be easier to detect from other high-z sources than one might expect extrapolating from local samples. If the strong emission holds true for additional high-z sources, then we will have identified further evidence that galaxies in the past underwent star-formation in a significantly different manner than they do today.

This work was supported by NSF grants AST-00736289 and AST-0722220, and NASA grant NNX10AM09H. We thank the CSO staff for their support of ZEUS operations and the anonymous reviewer for their helpful comments.

**Table 1**
Source Parameters

| Source Parameter | Units | SMMJ02399 | Ref. | The Cloverleaf | Ref. |
|---|---|---|---|---|---|
| RA | … | $02^h39^m51.9^s$ | … | $14^h15^m46.3^s$ | … |
| DEC | … | -01°35'59" | … | 11°29'44" | … |
| z |  | 2.8076 | Genzel et al. 2003 | 2.5579 | Barvainis et al. 1997 |
| $D_L$ | Gpc | 23.8 | … | 20.96 | … |
| Lensing Magnification[a] |  | 2.38 | I10b | 11 | Venturini & Solomon (2003); |
| [NII]122μm Line Flux [b] | $10^{-18}$ W m$^{-2}$ | 3.46±0.54 | this work | 3.03±0.29 | this work |
| [OIII] 88μm Line Flux [b] | $10^{-18}$ W m$^{-2}$ | 6.04±1.46 | F10 | … | … |
| $L_{FIR}$ | $L_\odot$ | 1.22E13 | Weiß et al. 2007 | 5.58E12 | Weiß et al. 2003 |

[a] All values are deboosted using the lensing-magnification factors indicated.
[b] Statistical uncertainties only. Flux calibration uncertainties are ~30%.

**Table 2**
Comparison between the Cloverleaf and M82

| Property | Units | The Cloverleaf[a] | Ref. | M82 | Ref. | Cloverleaf/ M82 Ratio |
|---|---|---|---|---|---|---|
| [NII] 122 μm Luminosity | ($L_\odot$) | 3.89E09 | this work | 8.36E06 | Colbert et al. 1999 | 470 |
| PAH 6.2 μm Luminosity | ($L_\odot$) | 1.93E10 | Lutz et al. 2007 | 1.72E08 | Wild et al. 1992 | 110 |
| PAH 7.7 μm Luminosity | ($L_\odot$) | 7.84E10 | Lutz et al. 2007 | 4.60E08 | Wild et al. 1992 | 165 |
| FIR Luminosity | ($L_\odot$) | 5.58E12 | Weiß et al. 2003 | 2.75E10 | [b] | 202 |
| Molecular Gas mass | ($M_\odot$) | 3.00E10 | Weiß et al. 2003 | 1.80E08 | Wild et al. 1992. | 170 |
| [NII]/ FIR | … | 0.0007 | … | 0.0003 | … | 2.3 |
| 6.2/ FIR | … | 0.0035 | … | 0.0062 | … | .55 |
| 7.7/ FIR | … | 0.014 | … | 0.017 | … | .84 |

[a]Cloverleaf values are intrinsic.
[b]Average of the values from Colbert et al.1999 ($3.2\times10^{10}\,L_\odot$); and Rice et al.1998 ($2.3\times10^{10}\,L_\odot$)

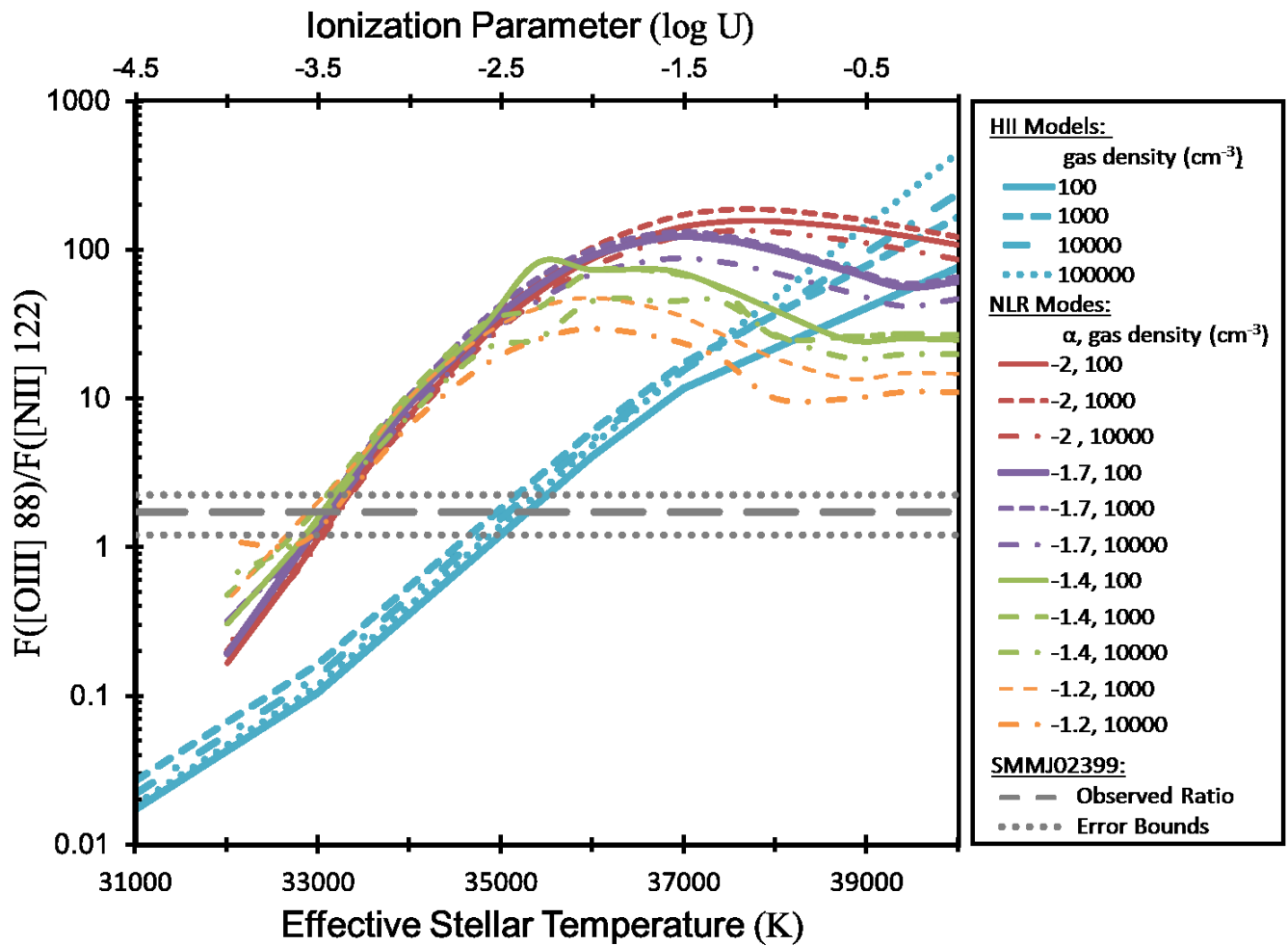

**Figure 1.** Observed [OIII]/[NII] line ratio from SMMJ02399 with error limits, and the expected line ratio as a function of the effective stellar temperature from the Rubin (1985) HII region models (lower axis) and ionization parameter of the NLR models from Groves et al. (2004; upper axis). The HII region models are plotted for several gas densities, while the NLR models are plotted for various power-law indices, α, and gas densities.

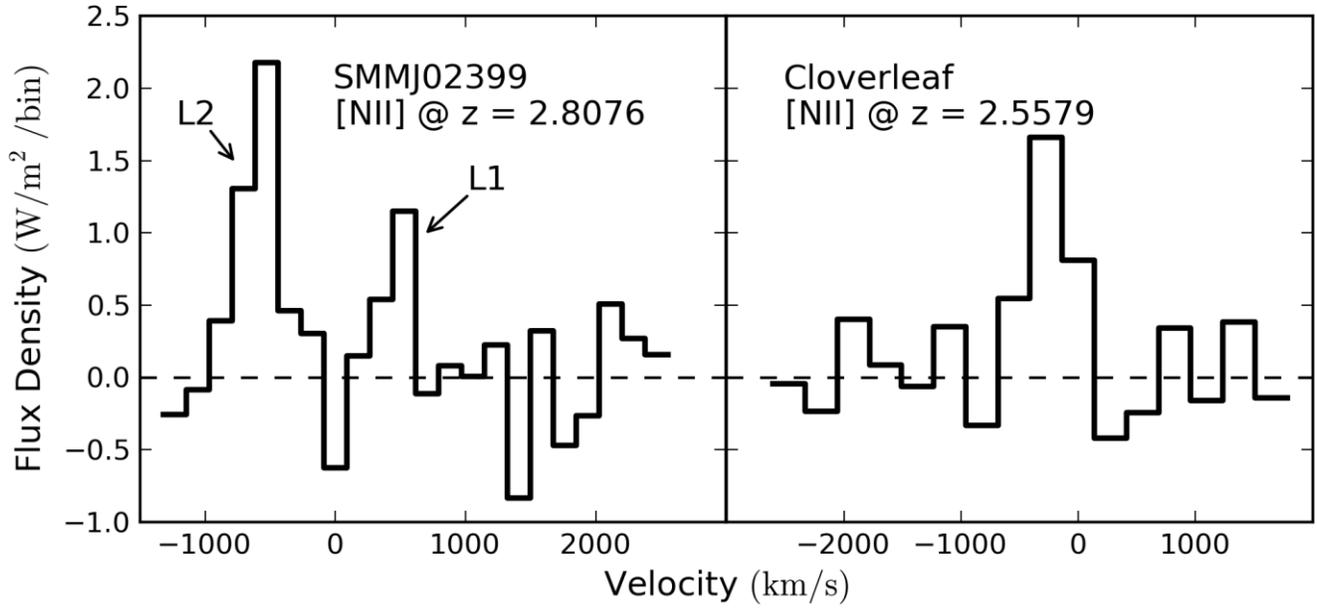

**Figure 2.** ZEUS/CSO detections of the [NII] 122 μm lines from SMMJ02399-0136 (left) and the Cloverleaf (right) plotted versus their rest-frame velocities. Spectral bins are ~ 1 resolution element, and equal to 200 km/s for SMMJ02399 and 300 km/sec for the Cloverleaf. The continuum emission has been subtracted.

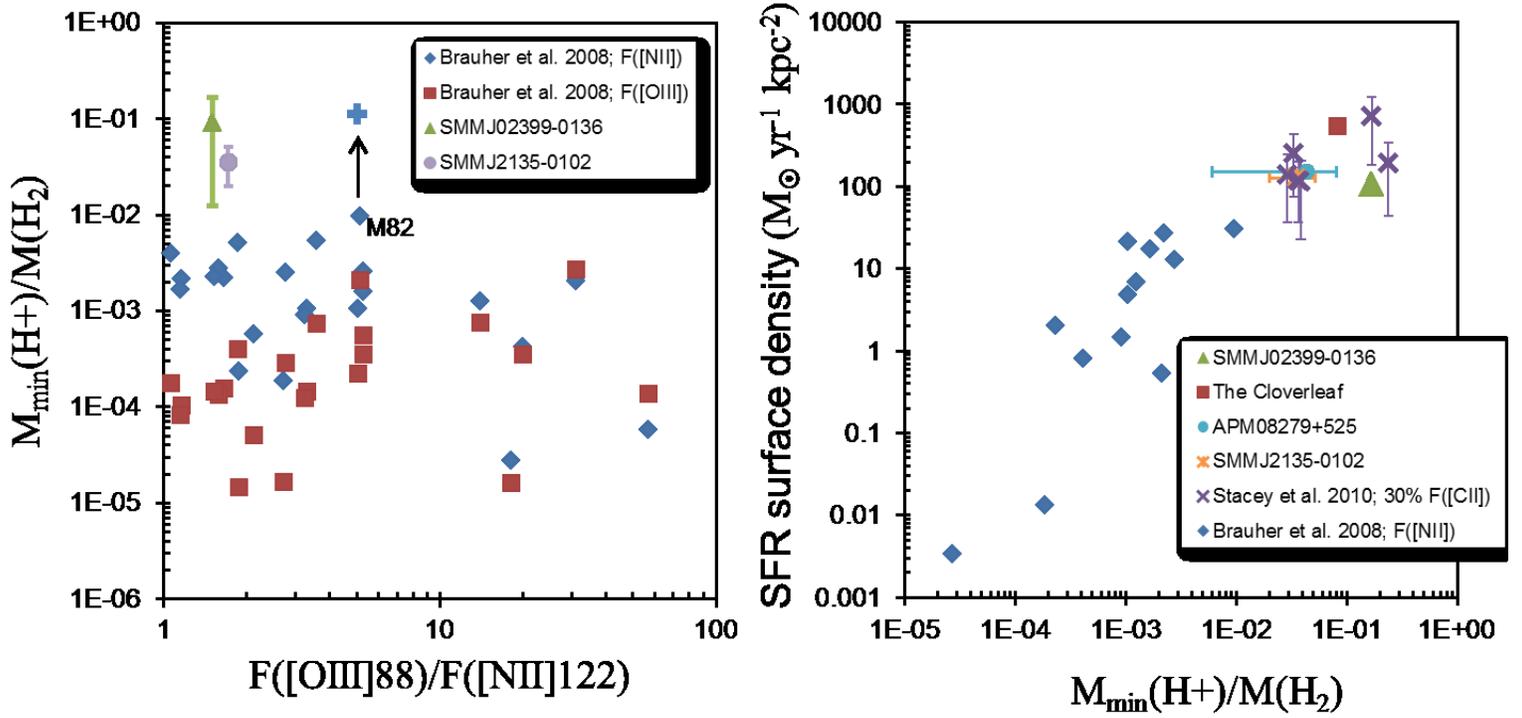

**Figure 3.** *(left)* $M_{min}(H^+)/M(H_2)$ fraction as determined using the [NII] and [OIII] lines versus the [OIII]/[NII] line ratio for local galaxies from Brauher et al. 2008, SMMJ02399, and line estimates for SMMJ2135-0102 from Ivison et al. 2010a. The error bars are the computed values and the symbols their average. Also shown is the mass fraction for M82 from Lord et al. 1996 (blue cross). *(right)* SFR surface density versus the $M_{min}(H^+)/M(H_2)$ fraction for SMMJ02399, the Cloverleaf, APM08279+5255, SMMJ2135-0102, a sample of high-z sources with [CII] 158 μm detections (Stacey et al. 2010), and the local sample. The SFR surface densities are from the literature or from their FIR luminosity using the Kennicutt (1998) scaling law.